# Emergence of Digital Twins

*Is this the march of reason?*

Dr Shoumen Datta, MIT Auto-ID Labs, Massachusetts Institute of Technology, Cambridge, MA

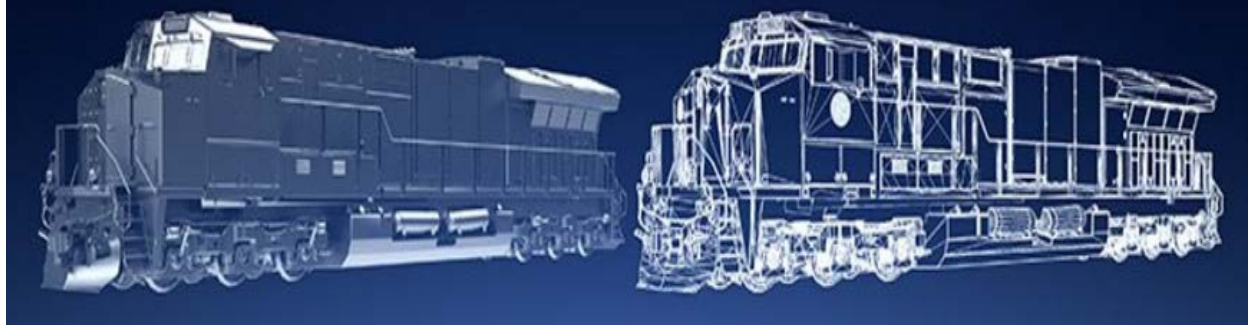

Multiple forms of digital transformation are imminent. Digital Twins represent one concept. It is gaining momentum because it may offer real-time transparency. Rapid diffusion of digital duplicates faces hurdles due to lack of semantic interoperability between architectures, standards and ontologies. The technologies necessary for automated discovery are in short supply. Progression of the field depends on convergence of information technology, operational technology and protocol-agnostic telecommunications. Making sense of the data, ability to curate data and perform data analytics at the edge (or mist rather than in the fog or cloud) is key to value. Delivering engines to the edge are crucial for analytics at the edge when latency is critical. The confluence of these and other factors may chart the future path for Digital Twins. The number of unknown unknowns and the known unknowns in this process makes it imperative to create global infrastructures and organize groups to pursue the development of fundamental building blocks and new ideas through research.

CONTENTS

# 1 INTRODUCTION

*Being Digital* (Nicholas Negroponte, 1996) and *When Things Start To Think* (Neil Gershenfeld, 2000) introduced into the public psyche the potential emergence and the rise of smart machines. About a decade later, Jeff Immelt of GE started to market these ideas in the *minds and machines* campaign claiming an imminent change to a future where self-organizing systems, sub-systems and multi-component subunits shall define the next generation of adaptive intelligent machines. When can we expect the *next generation* to start?

In 1513, the discovery of the isthmus at Panama by Vasco Núñez de Balboa triggered the idea of creating a trans-oceanic canal. Francisco Lopez de Gomara suggested (in his book, 1552) Panama, Nicaragua, Darien and Tehuantepec as choices for a canal. Not for another 300 years, not until the 19th century, would the canal building actually commence under the leadership of Ferdinand Marie Vicomte de Lesseps a French diplomat. Ferdinand de Lesseps (19 Nov 1805 to 7 Dec 1894) could not complete the Panama Canal and did not live to see the successful completion of the Panama Canal in 1914 by the US Army Corps of Engineers.[1]

Creating intelligent adaptive machines faces a similar uphill battle. Our optimism is not unfounded but it may be burdened by the dead weight of old technology. Paving the path for new theories, new concepts and new forms of connectivity in engineering design of future systems may lead to intelligent (?) machines. An even greater challenge may be understanding cognition in systems due to our poor grasp on intelligence[2] (AI).

Companies afraid to delve deeper are retrofitting existing machines to designate them as *connected*. Attaching sensors to collect data makes them *smart*. Workflow on steroids is the *intelligence* in analytics. Others are collecting and feeding big (volume) data sets to existing software systems and *voila* cognitive software systems emerge! The tapestry of buzz words and patch-work of programs are introducing glaring gaps, generating errors, callous disregard for physical safety and inept approach to cybersecurity[3] in general.

One reason for the confusion, perhaps, is our general inability to ask correct questions. These are some of the questions from the field. What machines, devices and systems may be built with the tools and technologies at hand? How should we build and use them? Do we really want to just connect everything to collect big volume of data? Is it really all about data? What is data curation? How can we teach smart machines to achieve specific goals? Are these the correct questions to ask? Are these questions worth answering?

The debate rages on about answers. These and other related questions may find some answers hidden in bio-inspired design principles. Progress in bio-MEMS, bio-NEMS and molecular machines[4] coupled with biological mimicry and cybernetics[5] are elements which may (?) converge with AI[6] in an over-arching strategic[7] plan. Integrating that plan in engineering design may be the Holy Grail. The command, control and coordination of bio-inspired engineering design requires hardware-software synchronization in the principle of the design.

Time-synchronized hardware-software integration is one hallmark of cyber-physical systems[8] (CPS) which is the foundation of embedded[9] systems. The concept of digital twins may have been suggested by NASA. Time guarantee (concurrence) in embedded systems plays a critical role in aero/astronautics. Advancing digital twins as an agenda for the industrial-information age must adopt practices[10] borrowed from CPS[11].

The current advocacy to advance the principles and pervasive practice of digital twins, from manufacturing to healthcare, calls for connectivity by design. Systems should be able to discover, inherit, evaluate and share intelligence across different sub-systems or sub-components. We should be able to monitor, analyze, control at the sub-unit level in real-time ([12]sensors, actuation) and visualize operations not only at the system level but the ecosystem[13]. The latter is perhaps the glue that may accelerate the future progress of digitalization.

## 2.1     SIGNAL VS NOISE – IOT VS DIGITAL TRANSFORMATION

The term IoT may have been coined at the MIT Auto ID Center (1999), but the past, present and future[14] concepts of IoT have been brewing for almost a century. A few milestones include Isaac Asimov's "Sally" the fictional autonomous car[15], Herbert Simon's seminal paper[16] ("talk to the computer"), Hiroshi Ishii's idea of "Tangible Bits" (People, Bits and Atoms[17]), Mark Weiser's paper[18] "Activating Everyday Objects" as well as the 1991 article in Scientific American[19] and the vision[20] of the "networked physical world" by Sanjay Sarma[21], David Brock and Kevin Ashton (2001). The IoT roadmap[22] promises to be even more dynamic in the future and scholarly discussions, including one by Alain Louchez[23] clearly outlines the layers of influence.

IoT is a digital-by-design metaphor and a paradigm for ubiquitous connectivity. The value proposition rests on proper use of the plethora of tools and technologies that must converge to make sense of the data. The hypothetical transparency is of little use without the data of things, if we wish to profit from IoT applications. On the other hand, digital transformation is a cacophony of ideas open to innovation from wireless systems[24] as well as broadband communication[25] and the advent of 5G which may enable time critical[26] operations. The latter, if combined with 8K[27] visualization, may catalyze robotic surgery. Masses may benefit from standard surgical procedures such as laparoscopic cholecystectomy, appendicitis and phacoemulsification (cataracts).

CNC machines, ERP, Web 2.0, fixed-task robots are examples of waves of digital transformation in business. The 2012 proposal[28] from Sanjay Sarma of MIT Auto ID Labs to pursue a **Cloud of Things** initiative resonated globally and the concept was promoted by others (Finland[29], France[30] and South Korea[31], to name a few). The next wave appears to be the transition from manufacturing products (as items to be sold) to the creation of a service ecosystem around the product to sell the service as a pay-per-use model. Digital transformation includes establishing a digital leash to monitor, promote, connect, track and trace in order to monetize every point of contact in the relationship (digital CRM) not only once (sales of product) but over the life time of the product. Hence, product lifecycle management evolves to digital PLM with quality of service (not product delivery) as the KPI and monetization tool. Quality of service (QoS) emerges as the *readiness* metric to gauge customer satisfaction. If QoS metrics are maintained by the provider(s) or manufacturer(s), then the client or customer is expected to pay for the QoS level associated with the product-service per contractual agreement.

In instances where the product is not an object (eg, teleco provider) the business models are inextricably linked to "outcomes" the customer expects. Monetization of digital transformation from an outcome-based model is complex due to the ecosystem of players and alliances. It is not easy to optimize and arrive at the point of convergence to deliver the outcome as a seamless function which involves an end-to-end value chain.

Even more complex is the task associated with monitoring each instance of engagement for micro-revenue and its disbursement. We need to track each instance and maintain a record of connectivity in an irrefutable evidence log (eg blockchain[32]). The latter may act as a digital ledger[33] to validate fractional micro-payments due from each point of contact (PoC). The digital id of the service delivered at the PoC identifies the member of the supply chain providing the unit of service at that specific instance. The latter is a part of the sum of services in the portfolio that defines the service and QoS which the customer expects. The customer pays for the final outcome (value in the value chain). The sum of the parts must be delivered before the value perishes. The duration of that value may be widely divergent (compare retail vegetables to predicting risk of diabetes).

Synthesis of the parts to act as a seamless function is the challenge. Who will build the parts of the platform which will be sufficiently open and interoperable to connect with the innumerable end points on the edge? Who will build the blocks to represent the digital functions? Who will build the blocks for the blockchains?

2.2   DIGITAL TWINS

Scenario -
Schlumberger is monitoring a drill-head in operation on a drilling platform in Outer Hebrides to determine the MTBF (mean time between failure) metric and trigger replacement to prevent work stoppage on the rig.

The camera at the tip of the drill-head and drill-head (drill-case) associated sensors (vibration, temperature, gyroscope, accelerometer) transmits (wired, wireless) video, audio and other data which must be analyzed as close to real-time as possible with respect to object identification, precision geolocation and process linkage. AI (?) analytics updates MTBF metrics. Depending on MTBF range (80%, 90%) as decided by business logic (when to replace) the drill-head spare parts supply chain (service, fulfillment) must be connected to auto-trigger the "head" when the MTBF range is reached. Purchase orders [supplier(s)] are followed by transport and logistics for delivery and workforce scheduling to execute the replacement prior to breakage (payment contracts, invoices and accounts payable are other points of connectivity). Data about the drill-head and lag time for each process/operation is captured by the operations management team, at a remote location, for future aggregate studies or collective evaluations. Can we visualize this entire end-2-end process as outlined?

In our current *modus operandi* this operation involves a plethora of operational silos (drilling operation, mechanical engineering, systems, supply chain, finance, human resources), software (connectivity between different locations, cloud infrastructure, cybersecurity) and hardware (not only the spare parts and drill-head but also the computational hardware/servers at different locations).

In its simplest form, the concept of DIGITAL TWIN posits that the flow of data, process and decision (as outlined above in this hypothetical scenario) is captured in a software ***avatar***[34] that mimics the operation.

The "twin" is the "digital" transformation which can be visualized by an analyst or manager most likely on a mobile device (phone, iSkin) in a manner that is location agnostic. Drilling down on a schematic illustration with the word "drill-head" links to the live video-feed from the drill-head camera which opens up on a new GUI (tab or window). Data fields (attributes, characteristics) related to the drill in operation (eg pressure, torque, depth, temperature, rotations/sec) is visible when one clicks on the icon for the drill. A plot showing the data approaching the MTBF metric may be instantiated using a command (such as "plot data") to show how the live data (from sensors) is feeding the dynamic plot showing the characteristics of the drill-head and the rate at which it is approaching the MTBF range set by the system (prescriptive and/or predictive values).

It may allow for a "what if" analysis if the analyst viewing the Digital Twin wishes to change the MTBF range and explore how the downstream processes may change (see principles of http://senseable.mit.edu/). The digital twin for the supply chain should spring into action showing delivery lag times from different suppliers and cost of normal vs expedited delivery. The material science attributes of the alloy used in manufacturing the drill-head should be visible. The analyst may use an *ad hoc* selection process and identify a new vendor. Can the system trigger process workflow to alert the people (roles) along the way to clear the requisition and generation of purchase order for the new supplier? Can it auto-verify the new supplier to check credentials, inventory, cost, transportation scheduling, quality of service reports and customer reviews of prior contracts?

We are still on the mobile device or laptop with the Digital Twin app. We have watched the drill-head in action and keeping an eye on a small window which shows the real-time data/analytics approaching MTBF. In the same field but on a different app we identified a new supplier claiming to custom-design 3D printed-on-demand drill-heads with precision fit (think 3D printed prosthetics, hip joints). To improve the fit, the new supplier in Tampere (Finland) downloads the video feed (from the cloud) of the drill-head operating in Outer Hebrides. The manager monitoring the end-to-end chain [a] selects the team of engineers who will replace the 3D printed drill using a HR menu which lists skill sets, proficiencies and years of expertise by category [b] pre-sets the command on the digital twin to actuate the replacement supply chain process when MTBF is 72% because fulfillment takes 21 days and by then the MTBF is predicted to reach 85% (code red – replace).

Each sub-unit provider must collaborate and synchronize (systems, standards, semantic interoperability) their role in the operation and the representation of their function in the digital twin model in real-time. In addition, the plethora of system providers, suppliers, third party software, analytics, cloud storage and hardware component manufacturers – all – expects to be paid for the "outcome" desired by the company.

The design of content and connectivity of such vast system of systems calls for entirely new principles of model-based systems engineering which will integrate global standards to anchor architectures[35] responsible to drive the digital by design paradigm. New models must inculcate the IoT digital-by-design metaphor with respect to connectivity by design, interoperability between standards by design and security by design. The silos of OT, IT and telecommunications must converge to create this new digital-by-design paradigm shift. In this forthcoming shift, objects and things may not be baptized after birth to follow the digital persuasion but will be born free of the analog baggage and will not need a path to digital transformation because they will be *born digital*.

### 2.2.1 CONFIGURING DIGITAL TWINS: CREATING THE BLOCKS - THE BLOCKCHAIN PARADIGM?

Lessons from cyber-physical systems (CPS) with respect to operational time synchronization may be key for certain forms of architecture for digital twins. Without open repositories, the process of creating (building) digital twins and the adoption of digital twins may be restricted to an industrial oligopoly. The vast majority of users cannot deploy an army of engineers to create custom digital twins for their exclusive experiments.

Rapid diffusion of digital twins calls for open source entity level models of sub-components (units). Think of each SKU listed in a BOM (bill of materials) as a system made up of sub-systems. Next, imagine each sub-system with unit parts to serve as the "block" or base level unit which needs to be created (built). The "old world" notion may have stopped at the physical manifestation - the actual unit made of tangible materials.

In the era of digital twins, we will call on the *source*, that is, the CAD/CAM model owner of that unit, to create and contribute to a common repository (?) the software representation of the unit replete with the physics of the material and the engineering characteristics of its operational functionality. For example, the physics of the part will inherit natural laws which governs all entities. For example, if a spare part were to fall off a table (on this planet), it will fall ***down*** at a rate defined by the acceleration due to gravity of 9.8 m/s$^2$. The latter is an inherited[36] attribute from the laws of physics (characteristic which forms the base in a 'layer cake' model).

To the informed mind, it is clear we have encountered and entered the domain of semantics and ontologies.

The digital twins of the granular units (parts), to be useful, must be connected by their relationships to the relevant data feeds from sensors/gateways. These entity relationship models and parts (connectivity) must be accessible to managers or analysts who can drag and drop the parts from the repositories on a "sense" table (device GUI). The ontology of entity level relationship models for digital twins may borrow from bio-inspired principles. Elements[37] from disease models, for example, a bio-surveillance model, is shown below.

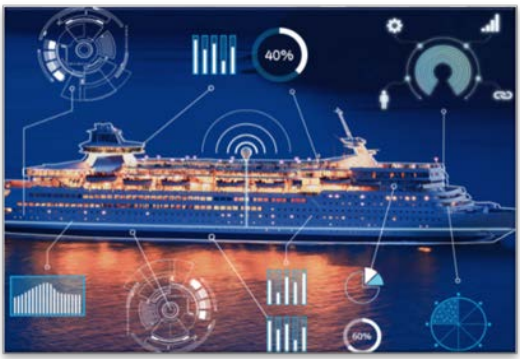 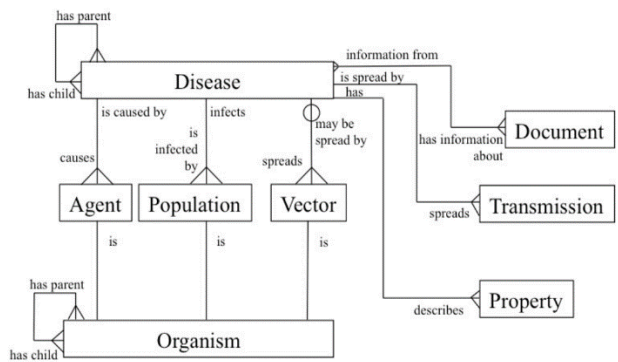

To the untrained eye the cartoons may not suggest the cryptic complexity that must form their foundation. These foundations are "layer cakes" (for example, TCP/IP) which must be able to communicate with other "layer cakes" (for example, semantic web[38]) built on other principles, concepts or ontological frameworks. Hence, it is imperative that we minimize the number of such architectures in order to fuel interoperability between these architectures (requires interoperability between standards, access to open data dictionaries).

The abstraction of the building blocks necessary for the digital twin movement may be similar, in principle, to the building blocks necessary to implement the use of blockchains, as a trusted ledger of connected instances. Who can we trust to build the blocks? This is a question of grave importance facing the practitioners of digital twins and blockchains. Both these concepts share homologies with IoT as a digital-by-design metaphor. The "block" in IoT may be the integrated platform synthesized from subunits or blocks containing data of things.

The rate limiting step, which defines the functionality of all of the above, is inextricably linked with and driven by the principles and practice of connectivity. In order to deliver value, connectivity must span a broad spectrum of dynamic ecosystems. Implementation of such connectivity must be protocol-agnostic, location agnostic and time sensitive (maximize transmission, minimize steps) with respect to the sense and response between the edge and the core. Since IoT is expected to connect trillions of things, scalability is a key enabler.

Have we encountered such "block" and "connectivity" concepts elsewhere? Perhaps the common answer may be in Agent[39] systems. Marvin Minsky's *brain connections* related abstraction[40] **"cube on cube"** illustrates this concept where each cube is a software Agent. It is relevant to this topic because each cube may be viewed as a "block" in the blockchain or a baseline 'unit' in the digital twin paradigm (digital copy of physical entity). The origin of the concept from software Agents, emphasizes the link to its semantics and ontology related roots.

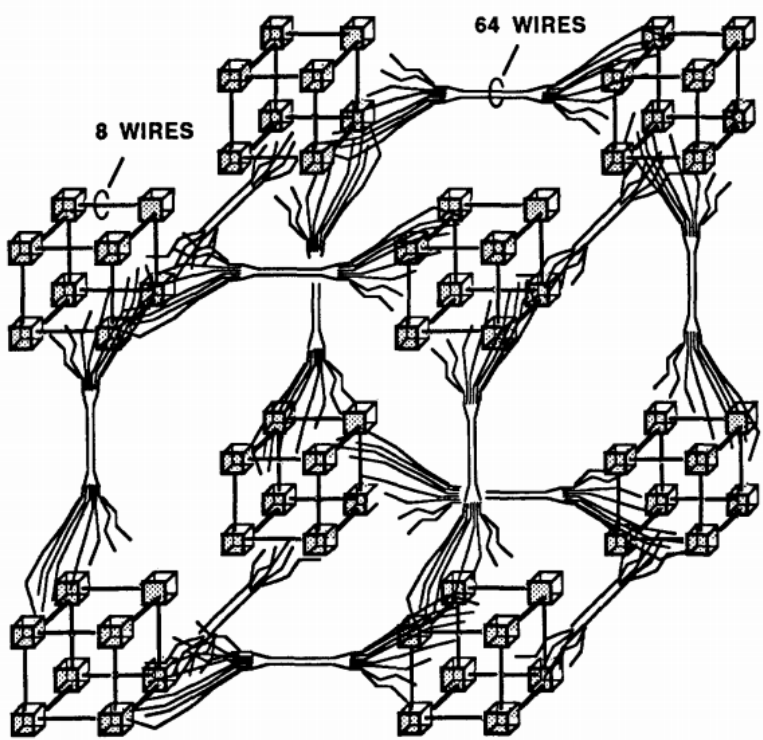

Illustration from page 315 (Appendix: Brain Connections) from *Society of Mind* by Marvin Minsky. MIT, 1985

The illustration (cube on cube) may render it easier to imagine how Minsky's abstraction and the principle of "blocks" may be useful to represent objects, data, process and decisions (outcomes). The blocks, if and when connected, may create or synthesize a variety of entities or networks[41] joined by common digital threads. For example, alignment of appropriate blocks can lead to creating platforms necessary for implementation of IoT. Parts and sub-units can be configured to create a digital twin of a machine (drill-head). Instances and units of transactions represented as blocks can constitute a digital ledger of events similar to financial blockchains.

For a scenario at hand, please consider the act of driving your automobile (if you can still use a gas guzzler with an internal combustion engine[42]) to the proximity of a dispenser in a gas station (petrol pump).

Your car recognizes "arrival" at the gas station, correlates with (perhaps) low reserve and unlocks the gas inlet. The petrol pump recognizes that your car is within the necessary proximity to the dispenser and recalls your choice for unleaded premium product. The nozzle from petrol dispenser discovers the gas inlet and commences fill-up when your inlet allows and confirms that the nozzle delivers petrol and not diesel. Once refueling completes you see a green icon on your dashboard. You receive a SMS indicating that completion of fueling triggered a financial transaction to match the amount of fuel. Your bank confirms payment over iSkin.

The convergence of IoT, digital twins and blockchain is evident (above). The ecosystem of enterprises, when dissociated by structure and associated by function in an operational sequence, presents a series of steps which can be sub-divided into "blocks" which are not only things or objects but also software Agents, unit of work, process, authentication, authorization, decisions, outliers, feedback, security, metrics, dependencies.[43]

**Who will build these blocks?**

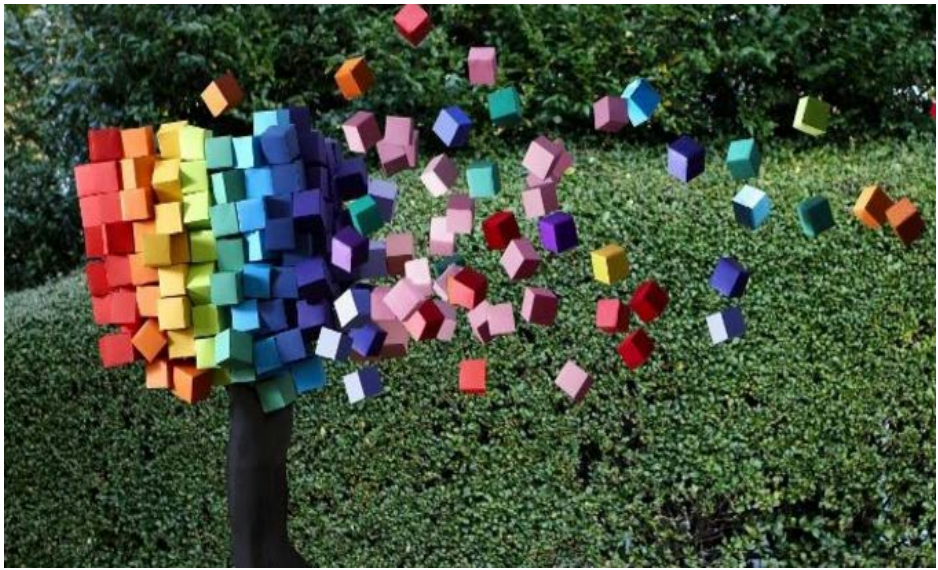

**Blocks** – a seamless operation represents an array of functions converging from diverse partner companies[44]

As it is with IoT, there will not be any one industry or company which may claim to be the front-runner. The fundamental building blocks for the domains spanning and overlapping IoT, digital twins and blockchains (not limited to financial transactions) are quintessential to the global economy. The idea of a distributed team or teams entrusted to architect these blocks may seem reasonable. But, the fractured state of the world and the intrinsic impact of natural language competencies on creating semantic dictionaries and ontological frameworks introduces socio-technical incongruencies. Hence, credible academic leadership of industry-government consortia in partnership with global organizations and standardization bodies may be an option.

If a few global alliances create the blocks and agree to establish the tools for interoperability, then we may anticipate a future global repository[45] for these digital blocks to accelerate global digital transformation. The ubiquitous need for principles and practice of connectivity[46] is salient to this discussion. The value expected from connectivity assumes the operation of multiple ecosystems which must converge to deliver the value. The table below suggests some of the layers and components necessary for this engineering ecosystem.

| 01 | I | Infrastructure | Scaffolds which includes energy, internet engineering, telco networks |
| 02 | T | Telecommunications | Backbone of connectivity which enables location agnostic operations |
| 03 | P | Protocol | Transaction triggered response operating agnostic of protocol-specificity |
| 04 | D | Discovery | Blocks/entities must find each other in order to communicate (think RDF) |
| 05 | C | Connectivity | Glue that enables digital transformation unless restricted by boundaries |
| 06 | S | Sense | Data acquired from points of interaction to understand status / attributes |
| 07 | R | Response | Analytics driven action/actuation based on integrating diverse knowledge |
| 08 | O | Operate | Outcome as pre-determined or change direction if influenced by factors |
| 09 | A | Adapt | Ability to remain dynamic and agile by recalibrating operations (eg SCM) |
| 10 | K | Knowledge | Learnings from operation (store/delete), dissemination, update analytics |

Local and global providers who supply the products and services germane to each layer (and several sub-layers within each layer) may not practice standard operating procedures (SOP). When volatility is the norm, it is wishful to expect SOP or expect groups in disparate parts of the world to conform. Hence, the task of interoperability and the ability to automate interoperability by "discovering" what is necessary in order to commence communication or cross-check various resources, becomes pivotal. It is a tool which is not yet available. Do we need this tool to discover and replenish the gaps in order for interoperability to commence?

Automating interoperability, at the least, may lead to auto-generation of APIs when interfaces "discover" that they cannot "cross-talk" between entities (models, data holders, tables, devices). It may trigger an automated mechanism to understand what needs to be understood between the systems and then obtains the "glue" (for example, creates a remote function call to source a "patch" from a repository) to enable interoperability.

If someone speaks to me in Hebrew, I must know that I am listening to Hebrew before I can use Google to communicate in Hebrew. If I had a tool to auto-detect languages then it could help trigger (use CNN/RNN[47]) Hebrew translation on my phone and empower my personal avatar or Siri or Cortana to guide my exchange.

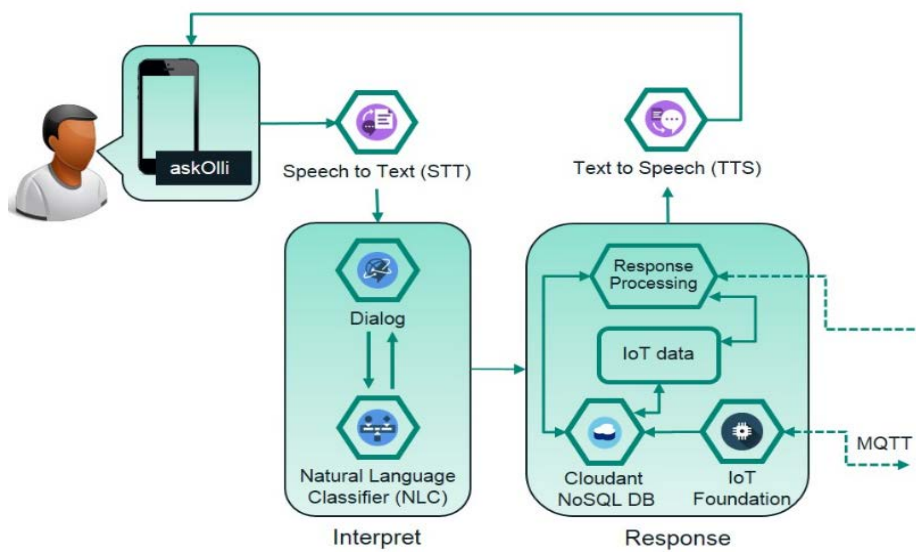

Discovery tools for auto-detection of attributes and characteristics between entities, Agents, models are a core part of the digital-by-design metaphor that IoT advocates. Research in connectivity may help develop this new generation of digital semantic sensors to sense (what we may not know) *what needs sensing*.

Therefore, claims for ubiquitous computing, first, must find tools for discovery. Media drum roll about trillions of "things" connected via IoT are delusional and hyperbole marketed by uninformed PR people. Unless objects can ***discover*** each other, they may not connect (assume that connectivity is protocol agnostic). Implementing the tools and technologies[48] central to "discovery" may accelerate digital transformation.

In the digital twin era, subunits will "discover" each other using embedded semantic properties to catalyze self-configuration (think *ad hoc* auto-configuration of mesh networks) to create the desired assembly (machine floor, medical devices attached to patient, turbines, water purification). Digital twins will inherit attributes of the physical components and physics of the system. Ontology-based semantic interoperability by design depends on entity level relationships. Distributed digital twins created by different sources may be able to communicate and form *swarms* to help us make even better decisions (one agent vs an agency or one ant vs a swarm of ants) employing the popular concepts of swarm intelligence.[49]

### 2.2.2 FRATERNAL TWINS: THE FIRST BORN – DIGITAL OR PHYSICAL?

A century ago (1916) a theory about freemartins[50] (the female of the heterosexual twins of cattle) generated interest about rare monozygotic twins[51] in cattle. This led to the discovery of Müllerian Inhibiting Substance[52] (MIS). Pioneering research[53] by Patricia Donahoe[54] is beginning to unravel the role and therapeutic potential of MIS. It appears that the human genetic program is inherently female[55] (which came first - male or female, the chicken or the egg). If the fetus was left to differentiate without MIS, fetal development of müllerian ducts will produce a female child – hence – the statement that human genetics is programmed to produce females. The fact that males exist is due to inhibition of the development of müllerian ducts by MIS and differentiation of the Wolffian ducts by fetal testosterone.

This very distant digression is intended to make the point that a fundamental plan, a base, exists in nature. The female plan is copied (duplicated) to produce the male, albeit, with modifications, catalyzed by MIS.

The concept of digital twins assumes we are creating a digital duplicate of the physical entity – the notion of being *born digital*. Can it be the other way around? In many instances in the industrial arena "things" cannot be created without an engineering plan and technical specs. The exception are humans (and animals) in case we combine patients and medical devices[56] to create digital twins, to monitor the physiological status.

A helicopter[57] may not be created as a physical entity unless we have a CAD/CAM software (digital) version and create a simulation (using differential equations) to test the operation (for example, rotation of the blades for lift-off). In recent models from major manufacturers (Boeing 787), the pilot is actually subservient to the simulated model in the auto-pilot. An image conjured by the latter generated the apocryphal statement that in the airplane of the future there may be only two living creatures in the cockpit. A pilot and a dog. The role of the dog is to stop the pilot from touching the controls. The role of the pilot is to feed the dog.

If we reverse the logic of the digital twins we have discussed thus far, one might propose the digital blueprint as the primordial layer and the physical entity to be the fraternal twin (limited mobility - think machines).

Therefore, the digital blueprint and the simulated models[58] which exists today, may be rapidly engineered with data feeds from the physical operation to approach the "live" concept of digital twins. In proposing this *modus operandi*, we move closer to the domain of cyber-physical systems[59]. Time dependencies create the need for time guaranteed software[60] which can understand the semantics of time, if time criticality is pivotal. For example, from (t=0) the decision to apply the brakes to the actual act of braking (t=n) to stop the car.

TEMPORARY CONCLUSION – CONFLUENCE OF SWARMS THROUGH FUSION – ALL ADVANTAGES ARE TEMPORARY

The digression about the conceptual see-saw about whose twin is it anyway is a thread of reasoning, not a barrier. It may make it easier to create the open repositories needed in the process of digital transformation. But, the road ahead for digital twins and digital transformation is still fraught with problems and also brimming with potential. Driving fusions (please see illustrations below) through collaborative ecosystems may be the only path to profit. There may not be a "winner takes it all" version in the digital twin economy.

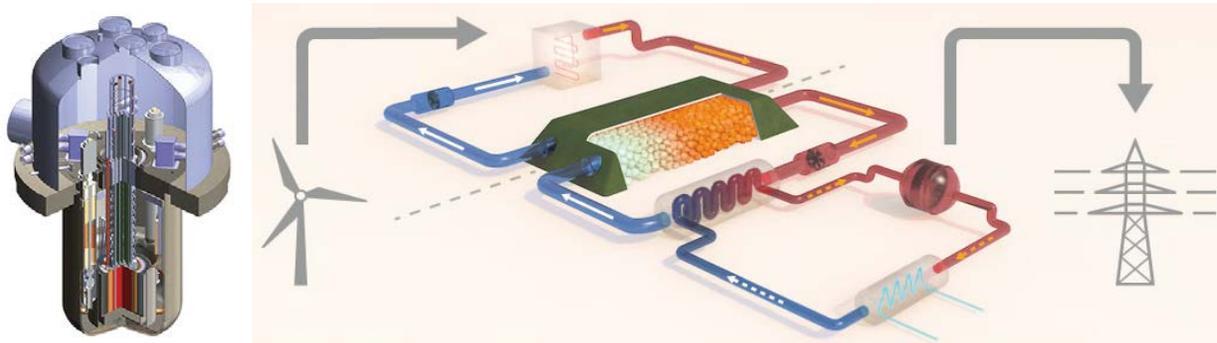

*Digital Transformation: Energy Equilibrium – Elusive Quest for the Digital Mitochondria? To maintain homeostasis of energy production, distribution and load balancing. Adapting to multiple sources and types of energy obtained from diverse producers (domestic, commercial) with variable end points (homes, roadside charger, factories, mobile delivery).*

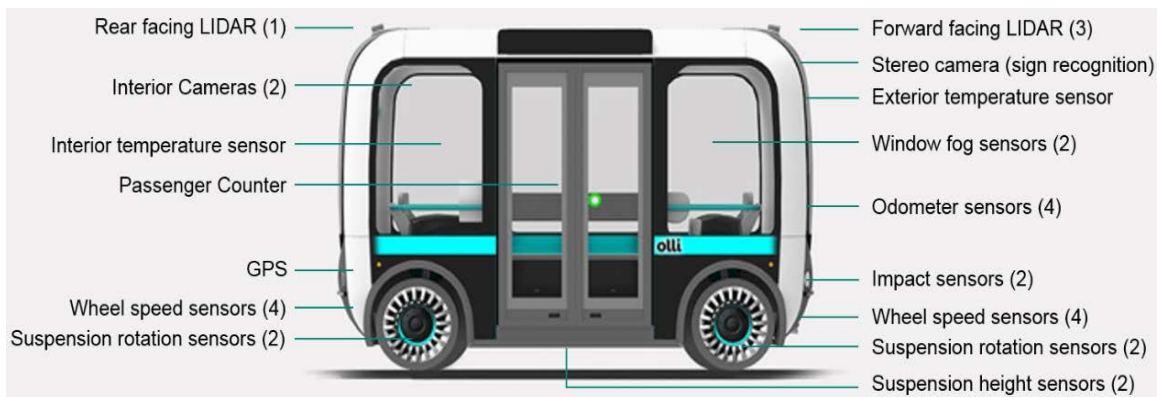

*Digital Transformation: The Transportation Alloy – Alliance of Autonomy, IoT, Telecommunications and 3D Printing* [61]

To profit from fusion in the digital-by-design era, collaborative efforts[62] may be one way forward. These examples of convergence (above) may be coupled with their operational digital twins or digital event duplication process. Information arbitrage from a wide cross-section of similar operations may provide a glimpse of patterns which were previously unobtainable due to the focus on one or few operations.

When we have a view not only of one operation (which is what one industry or one team wishes to monitor) but of a group of hundreds or thousands of such operations, we begin to understand patterns, predict faults, detect anomalies and use true "big" data and higher level metadata to feed other functions, such as data driven policy, security threats and intruder detection using pattern perturbation analytics.

Consider the cartoon (below, left) of a physical event and assume that we have a digital twin of that operation that an analyst or manager can remotely view to "see" or monitor the physical operation in progress.

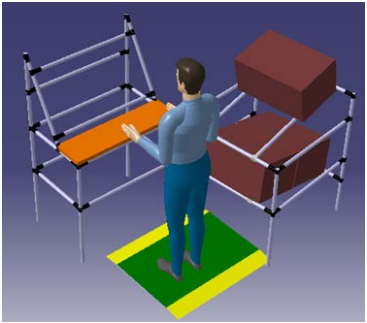 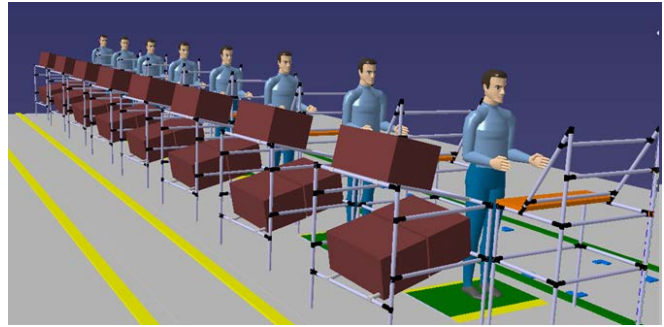

But, if the physical event (above, left) is not an isolated scenario, then digital duplication may generate (above, right) a form of digital transformation which may be representative of a digital swarm or flock[63]

This digital vision of aggregated events may generate big data and metadata from precision patterns can be extracted or extrapolated with respect to process, performance and profitability. While any one instance may not offer sufficient incisive insight, the application of the principles of swarm intelligence to hundreds of instances may provide wealth of information (not only data) which could add to the monetization[64] potential.

Detection of anomalies may improve predictive analytics (if a piece of equipment needs replacement) and errors or red herrings in the swarm may indicate security threat, breach or may elicit unusual activity alerts. The blockchain-like digital ledgers in the backbone of the digital twins may be useful in identifying the exact point of anomaly and the associated objects (including humans in the loop). If taken together with advances in hack-proof[65] code, this approach may add a new dimension to physical security and systems cybersecurity.

The swarm and flock approach if applied to the developing notion of "smart cities" may offer quite precise information from digital twin operations of scale-free networks in urban digital transformation. Monitoring digital duplicates of water valves in physical operation to control/regulate water waste, water security, water pollution. From an engineering systems point of view, the digital abstraction is applicable to city-level applications since cities are inter-dependent cascade of systems and networks[66] such as energy networks, traffic networks, sewer networks, communication networks, road networks, emergency response networks.

The vision of network convergence may be crippled and remain impotent without architectures which are resilient, fault tolerant and uses standards which are dynamic. But, interoperability between standards are rather difficult when competition fuels mistrust, spurs acrimony and short-term profits are the life-blood of the industry. Digital transformations calls for confluence of ideas beyond the horizon and new roads to reach the luminous summit. Investment in scientific[67] vision is often viewed with reservation, excessive caution, undue skepticism and even disdain. The latter is most unfortunate for the progress of civilization.

Digital twins, IoT, blockchains and swarm intelligence may re-define our imagination and future vision. *That* sense of the future requires businesses to re-think about ROI and profits, think about micro-payments and micro-revenue models but not at the expense of R&D or lowering the priority for innovation.

Industry must embrace innovation uncertainty and leaders must proactively support the call for creating structures necessary to pursue collaborative initiatives[68] through investment in workforce development, skills training, digital learning, education and research.